\documentstyle[12pt,cite]{article}
\catcode`\@=11
\renewcommand{\thefootnote}{\fnsymbol{footnote}}

\begin{document}

\begin{titlepage}

\hfill{DFPD/93/TH/62}

\vspace{0.2cm}

\hfill{hepth/9309096}

\vspace{0.4cm}

\centerline{{\large \bf QUANTUM RIEMANN SURFACES, 2D GRAVITY AND}}

\vspace{0.3cm}

\centerline{{\large \bf THE GEOMETRICAL ORIGIN OF MINIMAL
MODELS}\footnote[5]{Partly supported by the European Community Research
Programme `Gauge Theories, applied supersymmetry and quantum gravity',
contract SC1-CT92-0789}}

\vspace{0.6cm}

\centerline{\large{\sc Marco} {\sc Matone}\footnote{e-mail:
matone@padova.infn.it, mvxpd5::matone}}

\vspace{0.4cm}

\centerline{\it Department of Physics ``G. Galilei'' - Istituto Nazionale di
Fisica Nucleare}
\centerline{\it University of Padova}
\centerline{\it Via Marzolo, 8 - 35131 Padova, Italy}

\vspace{0.8cm}

\centerline{\large ABSTRACT}

\vspace{0.2cm}

Based on a recent paper by Takhtajan, we propose a formulation of 2D quantum
gravity whose basic object is the Liouville action on the Riemann sphere
$\Sigma_{0,m+n}$ with both parabolic and elliptic points. The identification
of the classical limit of the conformal Ward identity with the Fuchsian
projective connection on $\Sigma_{0,m+n}$ implies a relation between conformal
weights and ramification indices. This formulation works for arbitrary
$d$ and admits a standard representation only for $d\le 1$.
Furthermore, it turns out that the integerness of the ramification number
constrains $d=1-24/(n^2-1)$ that for $n=2m+1$ coincides with the unitary
minimal series of CFT.

\end{titlepage}

\newpage

\setcounter{footnote}{0}

\renewcommand{\thefootnote}{\arabic{footnote}}

\noindent
{\bf 1}. Recently in \cite{1} it has been developed an approach to
quantum Liouville theory based on the original proposal by Polyakov
\cite{p}.
The basic object in this theory is the `partition function of
$\Sigma_{0,n}$' with $\Sigma_{0,n}$ the Riemann sphere
punctured at $z_1,\ldots ,z_{n-1}$ and $z_n=\infty$
\begin{equation}
\langle \Sigma_{0,n}\rangle=
\int_{{\cal C}(\Sigma_{0,n})}{\cal D}\phi
e^{-{1\over 2\pi h}S^{(0,n)}(\phi)},\label{takpol1}\end{equation}
where the measure is defined with respect to the scalar product
$||\delta \phi||^2=\int_{\Sigma_{0,n}} e^\phi |\delta\phi|^2$,
and the integration is performed on the $\phi$'s such that $e^\phi$ be a
smooth metric on $\Sigma_{0,n}$ with asymptotic behaviour at the punctures
given by the Poincar\'e metric $e^{\phi_{cl}}$ (see (\ref{bhv2})).
The functional $S^{(0,n)}$ denotes the Liouville action
\begin{equation}
S^{(0,n)}(\phi)=\lim_{r\to 0}S^{(0,n)}_r(\phi)=
\lim_{r\to 0}\left[\int_{\Sigma_r}
\left(\partial_z\phi\partial_{\bar z}{\phi}+e^{ \phi}\right)+
2\pi (n {\log} r+2(n-2){\log}|{\log}r|)\right],
\label{32}\end{equation}
where $\Sigma_r=\Sigma_{0,n}\backslash\left(\bigcup_{i=1}^{n-1}
\{z||z-z_i|<r\}\cup\{z||z|>r^{-1}\}\right)$ and $z$ is the global
coordinate on $\Sigma_{0,n}$.
An important remark in \cite{1} is that by $SL(2,{\bf C})$-symmetry
one gets the {\it exact} result
\begin{equation}
\langle\Sigma_{0,3}\rangle={c\over |z_1-z_2|^{1/h}},\qquad
\Sigma_{0,3}={\bf C}\backslash\{z_1,z_2\},\qquad
c=\langle {\bf C}\backslash\{0,1\}\rangle,
\label{tpab}\end{equation}
which can be interpreted as correlation function of puncture operators
$e^{\phi/2h}$ of conformal weight $\Delta=\overline\Delta=1/2h$.
In \cite{1}, after fixing the standard normalization
$z_{n-2}=0,z_{n-1}=1,z_n=\infty$,
it is assumed that the theory defined by (\ref{takpol1}) satisfies the
conformal Ward identity
\begin{equation}
\langle T(z)\Sigma_{0,n}\rangle=\left[\sum_{i=1}^{n-1}{\Delta\over
(z-z_i)^2}+\sum_{i=1}^{n-3}\left({1\over z-z_i}+{z_i-1\over z}-{z_i\over
z-1}\right){\partial\over \partial {z_i}}\right]\langle \Sigma_{0,n}\rangle,
\label{takpol2}\end{equation}
where $T=(\phi_{zz}-{1\over 2}\phi_z^2)/h$ is the Liouville stress
tensor. Eq.(\ref{takpol2}) is verified at the tree level where
$\Delta_{cl}=1/2h=\Delta$ implying that
\begin{equation}
\Delta_{loops}=0.
\label{crucial}\end{equation}
Remarkably, in considering the tree level of (\ref{takpol2})
one uses \cite{1,p,ta}
the well-known relation between the accessory parameters and the
classical Liouville action \cite{0}
$c_i=-{1\over 2\pi} {\partial S_{cl}^{(0,n)}\over \partial z_i}$.
Expanding around the Poincar\'e metric $e^{\phi_{cl}}$, we obtain the
semiclassical
approximation\footnote{Note that in getting the second term in (\ref{takpol3})
one identifies $\left\{\phi_{cl}+\sum_k a_k\psi_k\big |
a_k\in {\bf R}\right\}$, where the $\psi_k$'s are the eigenfunctions of
$\Delta$, with the space ${\cal C}\left(\Sigma_{0,n}\right)$.}\cite{1}
\begin{equation}
\log\langle \Sigma_{0,n}\rangle=-{1\over 2\pi h}S_{cl}^{(0,n)}-
{1\over 2}\log\det (2\Delta+1)+{\cal O}(h),
\label{takpol3}\end{equation}
with $\Delta=e^{-\phi_{cl}}\partial_z\partial_{\bar z}$ the scalar Laplacian on
$\Sigma_{0,n}$. Eq.(\ref{takpol3}) implies \cite{1}
that the Ward identity works up to
one loop if $\Delta_{1\,loop}=0$ in agreement with (\ref{crucial}).

\vspace{0.3cm}

\noindent
{\bf 2}. It is natural to formulate a generalization of (\ref{takpol1})
in order to understand what is
the geometrical analogous of the correlators of Liouville vertices with
conformal dimension $\ne 1/2h$. To do this we first consider some facts about
Poincar\'e metric.

Near to an elliptic point the behaviour of the Poincar\'e metric
is\footnote{Here and in section 3 we consider the rescaled
field: $\phi\to \gamma \phi$, $\gamma=2h$.} \cite{kra0}
\begin{equation}
e^{\gamma\phi_{cl}}\sim{4q^2_k
r_k^{2q_k-2}\over\left(1-r_k^{2q_k}\right)^2},
\label{bhv3}\end{equation}
where $q_k^{-1}$ is the ramification index of $z_k$ and
$r_k=|z-z_k|$, $k=1,\ldots , n-1$, $r_n=|z|$. Taking the $q_k\to 0$
limit we get the parabolic singularity (puncture)
\begin{equation}
e^{\gamma\phi_{cl}}\sim {1\over r_k^{2}\log^2r_k}.
\label{bhv2}\end{equation}
Let $\Sigma_{h,m+n}$ be a Riemann surface of genus $h$ with $m$ elliptic
points $\{z_1,\ldots,z_m\}$ with ramification indices
$\{q_1^{-1},\ldots,q_m^{-1}\}$ and $n$ parabolic points ($p=m+n$).
Outside the elliptic points (the parabolic ones {\it do not belong} to
$\Sigma_{h,m+n}$) the Poincar\'e metric satisfies the Liouville equation
$R_{\gamma\phi_{cl}}=-1$, that is
$\partial_z\partial_{\bar z}\gamma\phi_{cl}=e^{\gamma\phi_{cl}}/2$.
Let $\overline \Sigma=\Sigma_{h,m}$ be the compactification of
$\Sigma_{h,m+n}$ (filling in the punctures).
The scalar curvature of $e^{\gamma\phi_{cl}}$ on $\Sigma_{h,m+n}$
$$
{R}_{\gamma\phi_{cl}}=-1+4\pi
e^{-\gamma\phi_{cl}}\sum_{k=1}^{m}(1-q_k)\delta^{(2)}(z-z_k),
$$
extends on $\overline \Sigma$ to $\overline{R}_{\gamma\phi_{cl}}=
{R}_{\gamma\phi_{cl}}+4\pi
e^{-\gamma\phi_{cl}}\sum_{k=m+1}^{m+n}\delta^{(2)}(z-z_k)$.
Therefore on $\overline \Sigma$
\begin{equation}
\partial_z\partial_{\bar z}\phi_{cl}={1\over 2\gamma}e^{\gamma\phi_{cl}}-
{2\pi\over \gamma}\left[\sum_{k=1}^{m}(1-q_k) \delta^{(2)}(z-z_k)
+\sum_{k=m+1}^{m+n}\delta^{(2)}(z-z_k)\right].
\label{mg0}\end{equation}
Note that Gauss-Bonnet formula implies
that $e^{\gamma\phi_{cl}}$ is not an admissible metric on
$\overline\Sigma$.

Let us now consider the $p$-point function in the standard approach to 2D
gravity
\begin{equation}
\langle \prod_{k=1}^p e^{\alpha_k\phi (z_k)}\rangle=\int_{{\cal C}(\Sigma_{h,0}
)}{\cal D} \phi e^{-{S^{(h,0)}\over 2\pi}}\prod_{k=1}^p e^{\alpha_k\phi (z_k)}.
\label{crlts6}\end{equation}
Here we do not care about the explicit form of the Liouville action. We only
assume $S^{(h,0)}$ be defined on a compact Riemann surface $\Sigma_{h,0}$,
and that the associated equation of motion be $\partial_z\partial_{\bar z}\phi=
{1\over 2\gamma}e^{\gamma\phi}$.
In the saddle-point approximation the
leading term reads
\begin{equation}
e^{-{S^{(h,0)}(\widetilde \phi)\over 2\pi}+\sum_k\alpha_k\widetilde\phi(z_k)},
\label{cllcl1}\end{equation}
where $\widetilde \phi$ satisfies the equation (note that
$\widetilde\phi\notin{\cal C}(\Sigma_{h,0})$)
\begin{equation}
\partial_z\partial_{\bar z}\widetilde \phi={e^{\gamma\widetilde \phi}\over 2
\gamma}-2\pi\sum_{k=1}^p\alpha_k\delta^{(2)}(z-z_k),
\label{mg1}\end{equation}
that for $\alpha_k={1-q_k\over \gamma}$,
coincides with eq.(\ref{mg0}). Eq.(\ref{mg1}) defines a $(1,1)$-differential
$e^{\gamma\widetilde \phi}$ which is not an admissible metric on
$\Sigma_{h,0}$. Nevertheless the previous discussion shows that eq.(\ref{mg1})
can be considered as the Liouville equation on the compactification $\overline
\Sigma=\Sigma_{h,m}$ of a Riemann surface $\Sigma_{h,m+n}$ with $n$-punctures
(where $n$ is the number of $\alpha_k$'s equal to $1/\gamma$) and $m$-elliptic
points where $e^{\gamma\widetilde\phi}$ coincides with the
Poincar\'e metric (for a discussion on admissible metrics
in this framework see \cite{grava}).
This investigation
suggests to extend the approach (\ref{takpol1})
by considering as basic object the Liouville action for Riemann surfaces
with\footnote{By abuse of language by `ramified points' we mean both
parabolic and elliptic points.} {\it ramified} points. In particular we will
still have the same classical limit as (\ref{crlts6}) but without the
constraint $\alpha_k=1/2h$. As a consequence we will get a purely
geometrical definition of conformal weight in Liouville gravity.
We recall that usually one defines conformal weights by {\it assuming}
validity of the
free field representation in order to perform the OPE.

Eqs.(\ref{bhv3},\ref{bhv2}) imply that the classical term
(\ref{cllcl1}) is divergent so that $e^{\alpha\phi}$ must be regularized.
The regularization is precisely the same that one considers
in defining the regularized Liouville action (\ref{32}). The crucial point is
that, as eq.(\ref{tpab}) shows, the regularization term fixes the
scaling properties of Liouville vertices. Similar aspects have been
discussed in \cite{grava}.

\vspace{0.3cm}

\noindent
{\bf 3}.
Here we shortly discuss the null vector equation arising in the CFT approach
to Liouville theory. The correctness of this approach needs to be
proved, nevertheless the following analysis will suggest a relationship
between conformal weights and ramification indices.

In \cite{bpz} it was pointed out that the uniformization equation for the
punctured sphere is related to the classical limit of the null vector equation
for the $V_{2,1}$ field $\psi$
$$
{\partial^2\psi(z)\over \partial z^2}+{\gamma^2\over 2} :
T(z)\psi(z):=0.
$$
In the CFT approach to Liouville theory one has
$T(z)={1\over 2}Q \phi_{zz}-{1\over 2}\phi_z^2$.
In \cite{gs} it has been proposed to compare the classical limit
$c_{Liouv}\to\infty$ of the decoupling equation for the null vectors in the
$V_{2,1}$ Verma module
\begin{equation}
\left({\partial^2\over \partial z^2} +{\gamma^2\over 2}
\sum_{i=1}^{n}{\Delta_i\over (z-z_i)^2}+{\gamma^2\over 2}
\sum_{i=1}^{n}{1\over (z-z_i)}{\partial\over \partial{z_i}}\right)
\big<V_{2,1}(z)\prod_{i=1}^{n}V_i(z_i)\big>=0,\label{gs}\end{equation}
$$\gamma = (Q-\sqrt{ Q^2-8})/2,\qquad c_{Liouv}=1+3Q^2,$$
with the uniformization equation
\begin{equation}
\left(\partial_z^2+{1\over 2} T^F(z)\right)\psi(z)=0,
\label{nffrmbis}\end{equation}
where $T^F=hT_{cl}$ is the Fuchsian connection
on the punctured Riemann sphere. This gives
\begin{equation}
\gamma^2\Delta_i={1\over 2}=\Delta_i^{(c)}, \qquad
i=1,\ldots,n,\label{cntrtr1}\end{equation}
where $\Delta_{p,q}={\alpha_{p,q}\over 2}(Q-\alpha_{p,q})$,
$\alpha_{p,q}={1-p\over 2}\gamma+{1-q\over \gamma}$,
and
$\Delta_i^{(c)}\equiv \Delta_{p,q}^{(c)}=\lim_{\gamma\to 0}\gamma^2
\Delta_{p,q}=(1-q^2)/2.$
Eq.(\ref{cntrtr1}) implies the constraint $\Delta_i=\Delta_{1,0},\,\forall i$,
so that $\big<V_{2,1}(z)\prod_{i=1}^{n}V_i(z_i)\big>=0$. Instead of `changing
uniformization' as proposed in \cite{gs}, we compare eq.(\ref{gs}) with the
uniformization equation
$\left(\partial_z^2+{1\over 2} T^{\{q_k\}}(z)\right)\psi(z)=0$,
where $T^{\{q_k\}}$ denotes the Fuchsian connection on the Riemann
sphere whose points $\{z_1,\ldots,z_{n-3},0,1,\infty\}$ have ramification
indices $\{q_1^{-1},\ldots,q_n^{-1}\}$. The important point is that now the
coefficient of the second order pole of $T^{\{q_k\}}$ at the elliptic points
is modified by a factor $1-q^2_k$ with respect to the parabolic case, that is
\begin{equation}
{1\over 2(z-z_k)^2}\;\longrightarrow\;{1-q^2_k\over 2(z-z_k)^2}.
\label{rmsts}\end{equation}

\vspace{0.3cm}

\noindent
{\bf 4}. By comparing eqs.(\ref{gs}-\ref{cntrtr1}) with eq.(\ref{rmsts}) we
have
\begin{equation}
\Delta_{cl}(q)=(1-q^2)/2h.\label{kdhd1}\end{equation}
Furthermore, the analysis in sect.2 shows that to a point of index
$q^{-1}$ we can associate a Liouville vertex of charge
\begin{equation}
\alpha=(1-q)/2h.
\label{aa}\end{equation}
In the following we will
show the correctness of eq.(\ref{kdhd1}) and will see that
$\Delta(q)=\Delta_{cl}(q)$.

Let us introduce the following `partition function of $\Sigma_{0,m+n}$'
\begin{equation}
\langle\Sigma_{0,m+n}\{q_k\}\rangle=
\int_{{\cal C}(\Sigma_{0,m+n})}{\cal D}\phi
e^{-{1\over 2\pi h}S^{(0,m+n)}(\phi)}.
\label{1a}\end{equation}
The functional $S^{(0,m+n)}$ denotes the Liouville action on $\Sigma_{0,m+n}$
whereas the domain of integration consists of smooth metrics on
$\Sigma_{0,m+n}$ with asymptotics given by (\ref{bhv3}) and (\ref{bhv2}) at
the points $\{z_1,\ldots,z_m\}$ and $\{z_{m+1},\ldots,z_p\}$ respectively.
For each ramified point the regularization term in $S^{(0,m+n)}$ reads
\begin{equation}
-2\pi\left((q-1)\log r + 2\log\left|{2q\over 1-r^{2q}}\right|\right).
\label{reg2}\end{equation}

Let $\Sigma_{0,1+2}$ be a Riemann sphere with a puncture at $z_3=\infty$ and
two elliptic points at $z_1,z_2$ with ramification numbers
$q_1^{-1}=q_2^{-1}=q^{-1}$. By $SL(2,{\bf C})$-symmetry we have
\begin{equation}
\langle \Sigma_{0,1+2}\rangle={c\over |z_1-z_2|^{1-q^2\over h}},
\label{jddjkpl}\end{equation}
so that we have the {\it exact} result
\begin{equation}
\Delta(q)={1-q^2\over 2h}.\label{hdqdlkm}\end{equation}
Let us set $z_{p-2}=0,z_{p-1}=1$ and $z_p=\infty$ ($q_p=0$) and denote by
$T^{(m+n)}(z)=(\phi_{zz}-{1\over 2}\phi_z^2)/h$
the stress tensor associated to (\ref{1a}).
Still in this case we assume the validity of the conformal Ward identity
\begin{equation}
\langle T^{(m+n)}(z)\Sigma_{0,m+n}\rangle=
\left[\sum_{i=1}^{p-1}{\Delta_i\over (z-z_i)^2}+
\sum_{i=1}^{p-3}\left({1\over z-z_i}+{z_i-1\over z}-{z_i\over z-1}\right)
{\partial\over \partial {z_i}}\right]\langle\Sigma_{0,m+n}\rangle,
\label{ptm13}\end{equation}
where
\begin{equation}
\langle T^{(m+n)}(z)\Sigma_{0,m+n}\rangle=
\int_{{\cal C}(\Sigma_{0,m+n})}{\cal D}\phi T^{(m+n)}(z)
e^{-{1\over 2\pi h}S^{(0,m+n)}(\phi)}.\label{takpol1bbb}\end{equation}
The tree level of (\ref{ptm13}) reads
\begin{equation}
T^{(m+n)}_{cl}(z)=\sum_{i=1}^{p-1}{1-q_i^2\over 2h(z-z_i)^2}-{1\over 2\pi h}
\sum_{i=1}^{p-3}\left({1\over z-z_i}+{z_i-1\over z}-{z_i\over z-1}\right)
{\partial S^{(0,m+n)}_{cl}\over\partial z_i}.
\label{ptm14}\end{equation}
By \cite{0} it follows that $-2\pi c_i=\partial_{z_i} S^{(0,m+n)}_{cl}$,
where now the $c_i$'s are the accessory parameters of $\Sigma_{0,m+n}$.
In this case the classical limit (\ref{ptm14}) reduces to the Fuchsian
projective connection $T^{\{q_k\}}$ (times $1/h$).
As before the semiclassical approximation of $\langle\Sigma_{0,m+n}\rangle$
implies that the Ward identity works up to one loop if
$\Delta_{loop}(q)=0$,
in agreement with (\ref{kdhd1}) and (\ref{hdqdlkm}).
The result in \cite{1} concerning the
evaluation of the Liouville central charge
extends to (\ref{1a}), that is
\begin{equation}
c_{Liouv}=1+{12\over h}.
\label{lccc}\end{equation}
In bosonic string theory
$h=12/(25-d)$, so that $c_{Liouv}=26-d$ and
\begin{equation}
\Delta_k\equiv\Delta(q_k)={(1-q^2_k)(25-d)\over 24}.
\label{ccwd3}\end{equation}

In order to interpret $\langle\Sigma_{0,m+n}\rangle$ in terms of
Liouville correlators we first recall that
in the DDK model \cite{ddk}
the modified Liouville action has the term
$\sim\int_{\Sigma}\sqrt{\widehat g}e^{\alpha\sigma}$ which is well-defined
only for $\Delta\left(e^{\alpha\sigma}\right)=1$.
Such a Liouville vertex can be represented by a ramified point.
However, by (\ref{ccwd3}), a necessary condition for the existence
of this representation is
\begin{equation}
\Delta(q)=1\longrightarrow q^2={1-d\over 25-d}.
\label{jfht}\end{equation}
On the other hand, since $0\le q^2\le 1$, it follows that the DDK model
has a geometrical counterpart only for
\begin{equation}
d\le 1.
\label{ojisdq}\end{equation}
This result furnishes a geometrical framework to consider the $d=1$ barrier
arising in the standard approach \cite{kpz,ddk} to 2D gravity coupled
to conformal matter.
Furthermore, since $q^{-1}\in {\bf N}$, we get
\begin{equation}
d=1-24/(n^2-1),\qquad n=q^{-1},
\label{ccri}\end{equation}
that for $n=2m+1$ is the unitary minimal series of CFT
$$
d=1-6/m(m+1).
$$
Note that by (\ref{jfht}) it follows that $d=1$ is related to a
puncture. In this case (\ref{reg2}) gets a $\log|\log r|$ term which is
reminiscent of the $\log$ correction to $\gamma_{str}$ for $d=1$.

By (\ref{aa}) and (\ref{jfht}) it follows that
\begin{equation}
\alpha={\sqrt{25-d}\left(\sqrt{25-d}-\sqrt{1-d}\right)\over 24},
\label{iuhe}\end{equation}
which should be compared with the rescaled value given in \cite{kpz,ddk}.
Note that positivity of $q$ implies not sign ambiguity in getting (\ref{iuhe}).
The relation (\ref{ccri}) between ramification index and central charge
is analogous to the relation arising
in the $k^{th}$-matrix model where $d=1-3(2k-3)^2/(2k-1)$. The value of
$k$ fixes the possible values of the deficit angle in the
triangularization.

\vspace{0.3cm}

\noindent
{\bf 5.}
We now discuss the origin of the $d=1$ barrier in the DDK model. To do
this we first consider the split
$$
{\cal D}g=d[\vec m]{\cal D}_gv^z {\cal D}_gv^{\bar z}{\cal D}_g\sigma
\det \nabla^z \det \nabla^{\bar z}.
$$
Since
$||v,v||^2_{g=e^\sigma\widehat g}=
\int_\Sigma \sqrt{\widehat g}{\widehat g_{ab}}e^{2\sigma}v^a v^b$, it
follows that ${\rm Vol}_g(Diff(\Sigma))$
depends on $\sigma$.
In critical string theory one usually {\it assumes}
that this dependence can be absorbed into ${\cal D}_g\sigma$ and then
drop the ${\cal D}_gv^z {\cal D}_gv^{\bar z}$ term.
However for $d\ne 26$ this procedure is not correct.
The question is to understand whether the DDK assumption in finding the
form of the Jacobian $J(\sigma,\widehat g)=e^{-S}$
still works when the term ${\cal D}_gv^z {\cal D}_gv^{\bar z}$ is included.
 A possibility to overcome this question
is to consider the partition function $Z=\int_{{\cal M}_h}{\cal Z}$ of
non critical strings
by investigating its properties by the point
of view of the theory of moduli
spaces of Riemann surfaces ${\cal M}_h$.

Of course ${\cal Z}$ must be a well-defined volume form on ${\cal M}_h$.
An important result about ${\cal M}_h$ is the Mumford isomorphism
$$
\lambda_n\cong \lambda_1^{c_n},\qquad c_n=6n^2-6n+1,
$$
where $\lambda_n=\det \,{\rm ind}\,\overline \partial_n$
are the determinant line bundles.
The fact that the metric  measure cannot depend on the background choice
implies that $c_{tot}=0$. It follows that
by the Mumford isomorphism ${\cal Z}$ is (essentially)
the modulo square of a section of the bundle
\begin{equation}
\Lambda =\prod_{k=1}^l \lambda_k^{d_k}, \qquad \qquad \sum_{k=1}^lc_k
d_k=0,
\label{trvlbndl}\end{equation}
where $-2c_jd_j$ is the central charge of the sector $j$.
In the Polyakov
string the matter and ghosts sectors have $d_1=-d/2$ and
$d_2=1$ respectively, thus (\ref{trvlbndl}) gives for the Liouville
sector $c_{Liouv}=26-d.$

Eq.(\ref{trvlbndl}) suggests to extend to the non critical case
the Belavin-Knizhnik conjecture \cite{bk}
(based on the GAGA principle \cite{serre}) concerning the
algebraic properties of multiloop amplitudes.

A way to represent CFT matter of central charge $d$
is to use a $b$-$c$ system of weight $n$, such that
$-2c_n=d$ \cite{bmtw}. Notice that,
since the maximum of $-2c_n$ is 1, this approach works  for $d\le 1$ only.
The model is exactly a CFT realization of the
Feigin-Fuchs approach where semi-infinite forms can be interpreted in
terms of $b$-$c$ system vacua.
Of course one can use the bosonized version
of the $b$-$c$ system which is equivalent to the Coulomb gas
approach.

For $d>1$ it is not possible to represent the conformal matter
by a $b$-$c$ system. In this case
one can consider the  $\beta$-$\gamma$
system of weight $n$ whose central charge is  $2c_n$. However
the representation of the $\beta$-$\gamma$ system in terms of free
fields is a long-standing problem which seems related to the
$d=1$ barrier.

Let us go back to eq.(\ref{trvlbndl}).
The question is to find the line bundle on ${\cal M}_h$ representing
 the Liouville sector.
The fact that $e^\sigma$ is positive definite suggests possible
mixing between Liouville, matter and ghost
sectors. In this context it is useful to recall that the Liouville
action defines a Hermitian metric on moduli space \cite{z}.

\end{document}